\documentstyle[12pt,epsfig]{article}
\begin{document}
\setlength{\parskip}{0.3cm}
\setlength{\baselineskip}{0.45cm}


%
\begin{titlepage}
\begin{flushright}
CERN-TH/99-324 \\
\end{flushright}
\vspace{1.5cm}
\begin{center}
\Large

{\bf Will relativistic heavy-ion
colliders\\
destroy our planet?}\\

\vspace{0.9cm}
\large
 Arnon Dar$^{\ast,\dagger}$, A. De R\'ujula$^\ast$ and Ulrich Heinz$^{\ast}$\\

\vspace*{0.5cm}
\normalsize
$^\ast$ Theory Division, CERN, CH-1211 Geneva 23, Switzerland \\
$^\dagger$ Department of Physics and
Space Research Institute,\\
Technion, Israel Institute of Technology, Haifa 32000, Israel\\
\vspace{1.5cm}
%
\large
{\bf Abstract} \\
\end{center}
\vspace*{-0.3cm}

\normalsize
\renewcommand
\baselinestretch 2
\noindent
Experiments at the Brookhaven National
Laboratory  will study collisions between gold nuclei at
unprecedented energies. The concern has been voiced that ``strangelets''--hypothetical products of these collisions--
may trigger the destruction of our planet.  We show how naturally
occurring heavy-ion collisions
can be used to derive a safe and stringent
upper bound on the risk incurred in running these experiments.

\vspace{5.5cm}

\noindent CERN-TH/99-324 \\
September 1999
\end{titlepage}
\newpage
\normalsize

\section{Introduction}

Experiments scheduled to start at the Brookhaven National
Laboratory (BNL) in the fall of
1999 will study heavy-ion collisions at record energies~\cite{RHIC}.
There has been a recent surge of concern regarding the
possibility that ``strangelets'' --hypothetical products of these collisions--
may initiate the destruction of our planet.
The trigger of this characteristically millenarian
concern may have been a comment by Frank Wilczek in
the July 1999 issue of Scientific American~\cite{FW},
comparing strangelets to ``ice-9'', a science-fiction
substance that would, on contact, freeze an ocean.
We derive a bound on the probability that the BNL experiments
may produce dangerous strangelets, not basing our
considerations on our theoretical understanding
of heavy-ion collisions,
but only on existing empirical
knowledge. Though our line
of argument is based on a succession of worst-case
choices, our results are appeasing.

Strangelets are hypothetical forms of nuclear
matter: single particles made of many $u$, $d$ and $s$
quarks. In putting together an ensemble of fermions
in their ground state,
it is advantageous to have as many different
particle types as possible, to circumvent the
exclusion principle. The substitution of a $u$ or a $d$
quark by an $s$ quark may be energetically favourable, in
spite of the penalty implied by the greater mass of the $s$.
On this basis Bodmer~\cite{Bo}
and Witten~\cite{EW}
suggested that strangelets,
like ordinary nuclear matter, may be stable.
There are tight upper limits
on the natural abundance of strangelets~\cite{DG,DR,Price,Lowder},
and reasons why
they may not have been produced in the early Universe~\cite{AF}.

Our understanding of the interactions between quarks
 is insufficient to decide with confidence
whether or not strangelets are stable forms of matter.
Estimates based on the MIT bag model~\cite{FJ,Madsen}
leave the   question open
for any mass (or baryon) number, $A$, between a single-digit
quantity and the value for neutron stars, $A\sim 1.7\times10^{57}$.
Stable nuclei have a charge-to-baryon-number ratio $Z/A\sim 1/2$.
Except for very small $s$-quark masses and for values of the chromodynamic forces between
quarks so large that the theory is no longer trustable, the estimated $Z/A$ for strangelets is positive~\cite{FJ},
but  much smaller than for nuclei. This is a direct
reflection of the borderline interplay
between the exclusion principle, which would favour identical
numbers of $u$, $d$ and $s$ constituents ($Z=0$), and the
mass excess of the $s$ quarks, which disfavours their constituency andresults in a positive $Z$.
A positive strangelet is not a threatening object. Exactly as an
atomic nucleus, it would gather an electron cloud and sit snugly
in whatever solid material it happens to find itself.

The recently headlined ``doomsday scenario''
 --whereby a strangelet would gather
atomic nuclei, become increasingly massive, fall to the
Earth's centre and accrete the whole planet-- requires the
theoretically unexpected existence of stable strangelets of negative charge.
Imagine that, for some unforeseen reason, there is a ``valley
of stability'' for negative strangelets. Suppose that, somehow,
such an object is produced in a laboratory high-energy
reaction and that it survives the collisions
that eventually bring it to rest in matter.
The negative strangelet would attract a positive nucleus
and may eat it. The resulting object may loose positive
charge and adjust its strangeness by electron capture
or positron $\beta$-decays.
The new strangelet may be negative again, and maintain an appetite for nuclei.
If its mass grows to some 0.3 ng ($A\sim 2\times 10^{14}$) it falls to the centre of the Earth~\cite{DG},
for its weight overcomes the structural
energy density of matter ($10^9$ erg cm$^{-3}$ or $\sim 0.1$ eV
per molecular bond).  At a mass above 1.5 ng, for a typical nuclear density,
the object becomes larger than an atom and the positron cloud that it has been
developing sits mainly inside the strangelet itself (for stable strangelets that
have grown this large, the sign of $Z$ is immaterial).
Even without the help of the Coulomb attraction, gravity and thermal
motion may then sustain the accreting
chain reaction until, perhaps, the whole planet
is digested, leaving behind a strangelet with
roughly the mass of the Earth and $\sim 100$ m radius.
The release of energy per nucleon should be of the order of several MeV
and, if the process is a run-away one, the planet would end
in a supernova-like catastrophe.

Experiments at the Relativistic Heavy Ion Collider (RHIC)
at BNL  should study Au--Au collisions
at a centre-of-mass (cms) energy $\sqrt{s}\sim 200\, A$
GeV $\sim 40$ TeV ($A\simeq 197$ for gold).
At the design luminosity of $2\times 10^{26}$ cm$^{-2}$ s$^{-1}$
and for the anticipated  six months per year of running, RHIC would
make some $2\times 10^{10}$ Au--Au collisions per year.
It would take RHIC 100 years to accumulate the statistics
gathered by NA50
 at CERN for Pb--Pb
collisions~\cite{NA50},
but this
fixed-target experiment was conducted at a smaller cms
energy: $\sqrt{s}\sim 17\, A$ GeV $\sim 3.5$ TeV
($A\simeq 207$ for lead). Since the Earth survived NA50
and all other Pb-Pb and Au-Au collision experiments
at CERN and BNL,
the ``BNL doomsday scenario'' must suppose that the formation
of ``killer'' strangelets occurs only at an energy above that of the
previous experiments, or that the difference between a collider
experiment (RHIC) and a fixed-target one (NA50) is not irrelevant.

In a proton--proton collision the probability of
producing heavy nuclei or antinuclei is utterly negligible,
as the energy required to make these relatively delicate
ensembles is far above their binding energy.
Strangelets would be, like atomic nuclei, fragile
objects that should be easy to disassemble.
In collisions between nuclei, in which the initial
baryon densities are high, the production of strangelets
may be favoured~\cite{Greiner,GRSK,GS}.
At the very high energies of RHIC at BNL, however,
it is very difficult to imagine how a strangelet
could be made and could survive~\cite{LH}.
But this question cannot be settled theoretically
with the tools at our disposal.
Moreover, it may not even be the right question.

{\bf We pose the question of whether one
can, on the basis of established facts, exclude beyond the
shadow of a doubt the ``BNL doomsday scenario''.}
This sort of question has arisen once and again as
new particle accelerators were built and operated.
The standard answer relies on a comparison of the
laboratory collisions with those that have occurred
in nature since the beginning of time. If the latter
have taken place in numbers enormously larger than those
envisaged in the lab, the probability of a catastrophic
outcome is correspondingly negligible. We shall see that
the most conventional reasoning --involving cosmic
rays impinging on stars and planets-- may contain a
potential loophole. But nature provides us with an alternative line of argument, which we use
to derive a fool-proof and stringent limit on the
potential danger of the BNL experiments.

\section{Traditional cosmic-ray limits}

Consider a strangelet made by a cosmic ray (CR)
in matter. Collisions of the RHIC-type
certainly occur in nature. Lead and gold nuclei
are similar. Lead is relatively
abundant in CRs, in interstellar gas,
or on the outskirts of celestial bodies without
a protective light-gas atmosphere, such as the Moon
or an asteroid. In a collision between a Pb CR
and a Pb nucleus at rest,
the CR energy equivalent to the RHIC cms
energy is $E\sim 4\times 10^3$ TeV.
This  is a modest energy by CR standards:
it is around the ``knee'' in the CR
spectrum~\cite{WB}.
The CR composition is measured directly
up to $\sim 100$ TeV and shows a relative abundance of heavy elements
which increases with energy. Extensive air-shower data
indicate that the trend continues at energies beyond the knee.
The CR flux is known,
from meteorite records, to have been steady for
billions of years. Reasoning along these lines
it is possible to deduce that, since the Moon
has not been destroyed by strangelet run-away
reactions, the probability of RHIC destroying
the Earth in five years of running is ``only''
of the order of
one in a thousand.  The
CR-induced
conversion of an asteroid into a large strangelet
--a ``killer asteroid'' that would in turn destroy the Sun as it
falls onto it-- leads to a stronger limit.
So does the production of a strangelet
in the collision of a Pb CR on an
interstellar Pb atom, with the strangelet continuing
its voyage into the Sun, and destroying it.

The argument regarding the Sun's survival
can be extended to the $\sim 10^{21}$ stars of the
visible Universe, which are not being destroyed
at a rate larger than that of supernova explosions.
The margin of safety is now astronomical. But, alas,
there is a potential flaw in the argument.

In RHIC, heavy-ion beams
of equal energy and opposite momenta will be made to collide
(the centre-of-mass system coincides in this case with
the laboratory system).
The hypothetical strangelets may be
produced with cms velocities $v$ that are {\it not} close to
the speed of light ($c=1$ in our units).
This small-velocity or ``central'' production is
completely contrary to the conventional
expectation~\cite{NA50,Greiner,GRSK,GS}
that strangelets ought to be mainly
made in the ``baryon-rich'' environment of the fast
forward- and backward-moving fragments of the
colliding nuclei. In the case of central production,
RHIC would be the first machine with the potential to make
strangelets nearly at rest in the laboratory.
In terms of the risk that we are discussing,
central production is the worst-case scenario, as we
proceed to explain.

Let $E_B\!\sim\! 7$ MeV be the typical nuclear binding energy
per nucleon. A nucleus with
kinetic energy per nucleon smaller than some 5 MeV
($v\!<\!v_{\rm crit}\sim\! 0.1$)
has a fair chance of surviving a collision with
another nucleus. A strangelet is also a form of nuclear matter,
and its binding energy per baryon (or per quark triplet)
cannot be much bigger than that of a nucleus:
the survival probability  in nuclear collisions cannot be very differentfor strangelets and for nuclei. This means that
a slow strangelet ($v\! <\! v_{\rm crit}$)
exiting a RHIC detector --and
colliding with iron nuclei in a magnet or with
concrete in the building-- should have a fair
probability of surviving without
being ruptured into potentially harmless
fragments. A slow and heavily ionizing
strangelet would come to rest after traversing
a column density of about 1 ${\rm g\,cm^{-2}}$,
at which point it might hypothetically start eating
nuclei. This is the gate onto the ``doomsday scenario''.

If only centrally produced in the cms ($v\! <\! v_{\rm crit}$),
``natural'' strangelets made by cosmic rays colliding
with stationary matter will be flying off with the cms
 Lorentz factor $\gamma=E/M$, which is of the
order of 100 for Pb--Pb collisions at the
RHIC-equivalent energy.
At this very high energy ($v\simeq 1$), the strangelet has
a very small probability of surviving a single nuclear
collision. Elastic (non-destructive) collisions
have small momentum and energy transfers.
Very many successive ones would be
necessary to bring the strangelet, unscathed,
to rest: the overall
survival probability ``exponentiates'' to a
truly tiny number. This may compensate for
the great number of strangelet-producing
collisions of cosmic rays with fixed targets
that have taken place in nature since
the dawn of time. It may also ``explain'' why NA50,
in spite of its large statistics, did
not trigger a cataclysm.
To avoid this conceivable loophole we look
for a natural imitation of an ion-collider facility.

\section{Heavy-ion collisions in space}

In-flight
collisions between cosmic rays are a rare but non-negligible
occurrence. In a fraction of these encounters the
centre-of-mass system moves sufficiently slowly for
the process to be similar to the ones studied at RHIC:
the flaw discussed in the previous section is avoided.
The risk incurred in running RHIC experiments can be
estimated by studying the putative effects of {\it slow} strangelets
made in CR--CR collisions. Rather than making a risk estimate, we
shall systematically impose exaggeratedly weak observational constraints,
thereby overestimating the danger.

Let $p$ be the probability to make
a slow strangelet in a single RHIC Au--Au collision. For
the planned running conditions, the number
of these particles made per year is
\begin{equation}
N=2\times 10^{10}\; p\,\; {\rm year}^{-1}\, ,
\label{prob}
\end{equation}
which will play the role of normalization.

For collisions whose cms velocity
$u$ or rapidity $y\!=\!\ln[(1+u)/(1-u)]/2$ are sufficiently small, therapidity distribution of the produced strangelets will be similar to
the $u=0$ rapidity distribution of an ion collider. We are interested
in collisions for which $u\! <\! v_{\rm crit}$, and we take as a reference
value $v_{\rm crit}\!\sim\! 0.1$, the velocity
below which we estimated a strangelet to be immune to nuclear
collisions. To satisfy this condition in the very high-energy collisions ofinterest, the momenta $p_i$ of the cosmic rays must
be nearly equal and nearly oppositely directed. Let
$\theta\!\sim\!2 p_T/E$ be
the (isotropically distributed) angle between $\vec{p}_1$ and
$\vec{p}_2$, with $p_T$ the total transverse momentum and
$E\!\sim\! E_1\!\sim\! E_2$. The fraction $f_\theta$ of collisions with cmstransverse velocity $v_T\!\sim\! p_T/(2 E)$
smaller than  $v_{\rm crit}$ is $f_\theta\sim 4 v_{\rm crit}^2$.
The condition for the cms longitudinal velocity to be smaller than
$v_{\rm crit}$ is that the ratio $E_1/E_2$ be in the range $1\pm v_{\rm crit}$. In the worst-case scenario in which strangelets
are only centrally produced,
we are exclusively interested in these nearly head-on collisions between CRs of nearly the same energy.

To obtain a lower limit on strangelet production in CR--CR collisions
we assume, conservatively, that strangelets are made only in
collisions between heavy nuclei (Pb--Pb or Au--Au) and
only above the RHIC energy $E_{\rm beam}\simeq 20$ TeV. The Pb abundance in CRs
of that energy is not directly measured, but the abundance and energy
spectrum of nuclei of the Fe group are~\cite{WB}.
At smaller energies,
the ratio of Pb-like nuclei to Fe nuclei
 is measured~\cite{PbFe}
to be $\sim\! 3\times 10^{-5}$; it is safe to adopt this
value at  higher energies, as the relative abundance
of the heavier elements increases with energy: they are more efficiently
accelerated and confined. We deduce that the flux $F$ and number
density $n$ of Pb in CRs are not less than:
\begin{equation}
{dF\over dE}={c\over 4\,\pi}\,{dn\over dE}
\simeq 5.3\times 10^{-11}\,\left[{E\over 1\;{\rm TeV}}\right]^{-2.6}\;({\rm cm^2\;s\;sr\;TeV})^{-1}\; .
\label{flux}
\end{equation}
We assume this locally measured flux to be representative of the CR
flux in the disk of galaxies such as ours.

The flux of Eq.\,(\ref{flux}) decreases very fast with $E$ and we are
restricting ourselves to CR collisions with $E_1\simeq E_2\!>\! E_{\rm beam}$. It is
therefore adequate to adopt an energy-independent
 strangelet production cross section
$p\, \sigma$, with $p$ the RHIC probability defined in Eq.\,(\ref{prob})
and $\sigma\sim 6.5\times 10^{-24}$ cm$^2$ the Pb--Pb nuclear cross section.
The rate per unit volume of strangelet production in the relevant
Pb--Pb CR collisions (whose cms is travelling with longitudinal
and transverse velocities
smaller than $v_{\rm crit}$) is:
\begin{equation}
R=2\,c\,p\,\sigma\,f_\theta\,\int_{E_{\rm beam}} dE_1\,
\int_{(1-v_{\rm crit})E_1}^{(1+v_{\rm crit})E_1} dE_2\,
{dn\over dE_1}\,{dn\over dE_2}\; .
\label{integral}
\end{equation}
The integral over $E_1$ converges so rapidly that it can be
extended to $E_1=\infty$.

Once produced, a charged strangelet with velocity
$v\! <\! v_{\rm crit}$ will be confined by a typical galactic magnetic field
$B\sim 3\,\mu$G to a region of size $3\times 10^{-11}\, A/|Z|$ kpc.
For $v=0.1$ and $|Z|=1$, interactions with ambient
hydrogen with an intestellar density of 1 atom per cm$^{3}$ bring the
particle to rest in a mere $5\times 10^6$ years. By galactic
standards, the strangelets stay put where they are born.
CR fluxes have been steady for billions
of years and were presumably larger some $T_0=10^{10}$ years ago,
when galaxies were young and the star formation rate (to which
the CR production rate should be proportional) was higher~\cite{S}.
We underestimate the accumulated number density
of strangelets in interstellar space as $n=R\, T_0$. Carrying out theintegrals in Eq.\,(\ref{integral}), we obtain:
\begin{equation}
n=R\, T_0\simeq 10^{-41}\, p\, \left[{v_{\rm crit}\over 0.1}\right]^3\,
\left[{20\,{\rm TeV}\over E_{\rm beam}}\right]^{3.2} \, {\rm cm}^{-3}\; ,
\label{density}
\end{equation}
where we have specified the energy dependence to facilitate comparison
with colliders other than RHIC.

A sufficiently large strangelet-production probability $p$ entails visible astrophysical consequences. How large can $p$ be?

\section{The fate of stars and planets}

Slow strangelets produced in CR--CR collisions come to rest and
accumulate in the material that is to become a star. We are interested in stars being born and dying at the current cosmological epoch.
At a typical interstellar density of 1 atom per cm$^{3}$, the material to
become a solar-mass star fills a volume $V\!\sim\! 10^{57}$ cm$^{3}$
(the average star is somewhat less massive than the Sun; we allow ourselves
a small degree of imprecision in various relevant parameters, since
the observational constraints we shall impose could be made very much
tighter).
The protostellar gas is concentrated, presumably by supernova shocks,
into a ``molecular cloud'', of density $\sim\! 10^3$ atoms per cm$^{3}$,
that collapses gravitationally. In all this process the strangelet
constituency, either at rest or magnetically confined, would follow
along with the ordinary matter and end up in the protostar.
The product $V\, n$, with $n$ as in Eq.\,(\ref{density}):
\begin{equation}
P_\star\equiv V\, R\, T_0 \sim 10^{16}\, p\, \left[{v_{\rm crit}\over 0.1}\right]^3\;
\left[{20\,{\rm TeV}\over E_{\rm beam}}\right]^{3.2}\; ,
\label{rate}
\end{equation}
is the probability for a solar-mass star to contain a strangelet
(or, if $P_\star\!>\!1$, the average number of strangelets it would contain).

In the consumption of a star by a strangelet
the energy release is of the order of the gravitational
binding energy $\sim\! GM^2/r$ of the strange remnant.
For a solar mass star and typical nuclear density, $r\!\sim\!10$ km
and the energy release is $\Delta E\!\sim\! 10^{53}$ erg,
two orders of magnitude bigger than the time-integrated
kinetic and visual energy of a supernova. A
star of less than a few solar masses would not become a black hole
that could potentially engulf all released energy. The late stages
in the conversion of a lighter
star into strange matter are presumably akin to a supernova
explosion. We discuss in turn a supernova-like signature and a putative
slower star consumption.

A typical galaxy contains $N_\star\!\sim\! 2.5\times 10^{10}$ stars,
currently dying as supernovae at a slower rate than
$R_{\rm SN}\!\sim\! 5$ per millenium~\cite{BT}.
The corresponding rate at which $N_\star$ stars are being destroyed
by strangelets is $R_{\rm destr}\!\sim\!N_\star\,P_\star/T_0=V\,R$, with
$P_\star$ as in Eq.\,(\ref{rate}). It would not be possible to have a 50\%
addition of a completely new type of (strangelet) supernova
without unacceptably
upsetting our understanding of this field. The condition
$R_{\rm destr}\!<\!R_{\rm SN}/2$ yields
\begin{equation}
p < 10^{-19}\, \left[{0.1\over
v_{\rm crit}}\right]^{3}\;
\left[{E_{\rm beam}\over 20\,{\rm TeV}}\right]^{3.2}\; .
\label{limit}
\end{equation}
Compare this result with the RHIC rate of Eq.\,(\ref{prob}). For the RHIC
beam energy and the reference $v_{\rm crit}$, our
extremely conservative conclusion
is that it is safe to run RHIC for
500 million years. This is reassuring, but what if the
conversion of a star into strange matter occurs over a longer period
than the visual display of a supernova?

We should only be concerned about the destruction of the
Earth in less than $T_0\!\sim\! 10^{10}$ years, since by the time the
age of galaxies doubles,
the Sun will have become a red giant and engulfed our planet.
We do not attempt to estimate the time it takes a strangelet
to become large enough to sink to the centre of a planet or
a star, but
the rate per unit mass at which a strangelet would
ingurgitate the Earth is certainly inferior
to the corresponding rate for a star, for all the conceivably
relevant parameters (temperature, pressure, speed of sound,
gravitational free-fall time,...) favour a faster star rate.
We conclude that the time for a strangelet-contaminated star
to develop an Earth-mass strange core is smaller than the time
it would take a strangelet-contaminated Earth to be destroyed.
How much longer would it take for the rest of the star to be processed?

Even if strangelets are lighter (at fixed baryon number)
than nuclei, we do not expect $^{56}$Fe, say, to decay into a
strangelet containing $\sim\! 56$ $s$-quarks. The reason for this is thatthe states of intermediate strangeness may not be less massive
than Fe, and
the overall decay process is a $\sim\!56$th-order weak interaction. The
ominous scenario that we are discussing tacitly presumes that for
sufficiently large strangelets such a decay barrier does not exist,
and the weak first-order transitions which process ordinary into
strange matter ($u\, d\to s\, u$, $u\, e^-\to s\,\nu$, and
 $u\!\to\! s\,e^+\,\nu$) can occur unimpeded.  The $u$-excess constituency
provided by ordinary matter accreting into a strangelet would
then exponentially decay away with a time constant comparable
to, or faster than, that of neutron decay (ten minutes). The rate
of consumption of a star would be governed by the much slower
rate at which matter can accrete onto the core strangelet.
Let $m_p$, $n_p$ and $v_p$ be the proton mass, number density and thermal
velocity in the neighbourhood of the strangelet's surface.
We estimate the mass-accretion rate as:
\begin{equation}
{dM\over dt}\sim m_p\,n_p\,v_p\,S=m_p\,n_p\,v_p\,4\pi
\left(3\,M\over 4\pi\,\rho_s\right)^{2/3}\, ,
\label{eat}
\end{equation}
where $\rho_s$ is the strangelet's mass density and $S$
is its surface.
For the sake of guidance, adopt the conditions prevailing in the
center of the Sun: $m_p n_p\!\sim\!1\!$~kg/cm$^3$, $v_p\!\sim\!10^{-3}\,c$,$\rho_s\!\sim\!10^{39}\,m_p/$cm$^3$. The
result for a solar-mass star is then $t\!\sim\! 130$
years, negligible with respect to $T_0$. In worrying about the
Earth's survival until the Sun engulfs it,
we are therefore concerned with stars becoming strange in the same
span of time: $\tau\!<\!T_0$.

For $\tau$ shorter than $\sim\! 300$ years, a single strange star would have a luminosity
$\Delta E/\tau$ superior to the bolometric luminosity $L\!\sim\!10^{43}$
erg/s of a galaxy containing $N_\star$ stars;
this case is covered by our previous
considerations on supernovae. For  longer $\tau$, very conservatively,
we demand that the ensemble of strange stars in a galaxy be insufficientto overshine the normal stars: $P_\star\, N_\star\, \Delta E/\tau\!<\! L$.
The weakest condition is obtained for $\tau=T_0$, and it gives the
same numerical result as Eq.\,(\ref{limit}).

It could be agnostically argued that, since the process of accretion onto a strangelet
is surely difficult to model with confidence, our use of Eq.\,(\ref{eat})is suspicious, and we should only abstract from it the fact that the accretion times of
different objects may be in proportion to the cubic root of their masses, which
is the result for fixed $n_p\, v_p$. For a
solar-mass star the time would be $\sim\!100$ times longer than for
our planet. The last paragraph's argument, for $\tau\!=\!10^2\, T_0$, gives
a condition two orders of magnitude weaker than Eq.\,(\ref{limit}).
Comparing with the RHIC rate of Eq.\,(\ref{prob}) we would then deduce, in this most unnaturally pessimistic case,
that running the RHIC experiments for five million years is
still safe.

\section{ALICE at the LHC}

At the Large Hadron Collider (LHC) currently being built at CERN,
the experiment ALICE~\cite{ALICE}
will study Pb-Pb collisions at
$E_{\rm beam}\!\sim\! 600$ TeV, roughly 30 times higher than at RHIC.
At the LHC, the planned number of heavy-ion collisions per year is similar to the corresponding figure at RHIC. To analyze the LHC
case in the same spirit with which we
have studied RHIC, we have to raise the threshold energy
for strangelet production to the LHC energy, even though
this assumption was already ultra-conservative for RHIC. Raising
the threshold energy, and reusing Eq.\,(\ref{rate}),
we conclude that the safety margin for ALICE is a factor
$30^{3.2}\!\sim\! 5.3\times 10^4$ lower than it is for RHIC.
This means that, in discussing ALICE, it would presumably be
advisable to improve our very safe limits based on the fate
of stars and/or to develop considerations that rely more heavily
than ours on our understanding of heavy ion collisions.
For example, if one were to argue that, at a fixed energy per
nucleon, Fe-Fe collisions are as good or better than Pb-Pb
collisions
at making strangelets, the probability $P_\star$ in Eq.\,(\ref{rate})
would
increase by about 11 orders of magnitude, due to the smaller
equivalent CR energy per nucleus, and the much larger CR abundance of Fe. The safety margins we have derived would
improve by the same factor.

\section{Discussion and conclusions}

We have argued that the experiments at RHIC do not represent a threat
to our planet. But, is this ``beyond the shadow of a doubt''? Considerations
analogous to ours have been made for other questionably dangerous
physical possibilities, such as the production of black holes or the
trigger of a reaction whereby the vacuum in which we are would be
catastrophically
converted into a ``true'' vacuum of lower energy density~\cite{SC}.
In these cases
one is dealing with relatively simple theoretical constructs and one
can draw conclusions that are correspondingly uncontroversial. In thecase of strangelets, we are dealing with the properties of an
incompletely understood hypothetical form of nuclear matter. It is always possible to come
up with an ``ad hoc'' hypothesis and invalidate any arguments. In the case
at hand, it would suffice to assume that strangelets are stable
only for masses smaller  than
the mass of the Earth, so that the conversion process to strange
quark matter is eventually stopped. Even if all stars contained a
stable Earth-mass strange core, it would not be easy to tell.
To have the upper limit of strangelet stability at a mass comparable
to that of the Earth, it is necessary to tune
the parameters of the underlying theory to a relative precision $\epsilon$of the order of the ratio of
a typical nuclear binding energy to the rest energy of the Earth, $\epsilon\!\sim\!10^{-49}$.
The a priori probability for the parameters to be so fine-tuned is of order $\epsilon$.
This gives an idea of how exceedingly
ad hoc any hypothesis of this kind would have to be.

We conclude that,
beyond reasonable doubt,  heavy-ion
experiments at RHIC will not endanger our planet.

{\bf Acknowledgements}.
 We thank A. Cohen, B. Gavela, R.L. Jaffe, L. Maiani and M. Sher
for fruitful discussions.

\noindent
{\bf Note added.} After the completion of our manuscript we received an article
by W. Busza {\it et al.}, hep-ph/9910333, in which limits stronger
than ours are derived, with use of arguments based on heavy ion
collision theory.

\end{document}